 \documentclass[preprint]{revtex4}  
\usepackage{graphicx}
\usepackage{dcolumn}
\usepackage{bm}
\usepackage[usenames]{color}

\begin{document}

\title{Sub two-cycle soliton-effect pulse compression at 800\,nm in Photonic Crystal Fibers}

\author{Marco V. Tognetti and Helder M. Crespo}
\email{marco.tognetti@fc.up.pt}

\affiliation{CLOQ/Departamento de F\'isica, Faculdade de Ci$\hat{e}$ncias, Universidade do Porto, Rua do Campo Alegre 687, 4169-007 Porto, Portugal}

\begin{abstract}
The possibility of soliton self-compression of ultrashort laser
pulses down to the few-cycle regime in photonic crystal fibers is
numerically investigated. We show that efficient sub-two-cycle
temporal compression of nanojoule-level $800$\,nm pulses can be
achieved by employing short (typically $5$-mm-long) commercially
available photonic crystal fibers and pulse durations of around
$100$\,fs, regardless of initial linear chirp, and without the need
of additional dispersion compensation techniques. We envisage
applications in a new generation of compact and efficient sub-two
cycle laser pulse sources.
\end{abstract}


\maketitle 

\section{Introduction}

Over the last  twenty  years,  much  effort  has  been  done in
generating ultrafast laser pulses in the few-cycle  regime. These
pulses  are  behind numerous and important achievements in such
diverse fields as the  study  of ultrafast dynamics in matter,
extreme nonlinear optics, high-order  harmonic generation, and
attosecond physics\,\cite{Brabec,Kartner}.

Low-energy (nanojoule) sub-2-cycle pulses have  been  directly
produced  by state-of-the-art Ti:sapphire laser
oscillators\,\cite{Morgner,Ell}  where  the  required ultrabroad
bandwidths were achieved by means of double-chirped mirrors  that
provide high reflectivity bandwidth  and  tailored  dispersion
compensation over an extended spectral range. Higher  energy  pulses
(from  hundreds  of nanojoules up to  hundreds  of  microjoules)
have  also  been  produced  by passive and/or active phase control
of   broadband  laser  sources  and  of supercontinua  generated  in
Kerr  media.  These  include   ultra-broadband optical  parametric
amplifiers\,\cite{Baltuska},  spectrally  broadened   cavity-dumped
oscillators\,\cite{Baltuska1}, and gas-filled  hollow  fibers pumped
by  kHz  amplifiers\,\cite{Nisoli,Nisoli1} where active compression
of  the  generated  broadband  spectra  has resulted in the shortest
($3.4$ fs) light pulses to date\,\cite{Schenkel,Yamane}.

However, in all of the above techniques, the  possibility  of
reaching  the few-cycle  regime  relies  on  sophisticated
intracavity   or   extracavity dispersion compensation schemes, with
the latter usually requiring complex electronically-controlled
feedback systems.

A different approach to pulse compression relies on the concept of
soliton-effect compression of laser pulses, which dates back to the
1980s\,\cite{Agrawal} and has been used to successfully compress
low-energy picosecond  pulses  to durations as short as $18$\,fs\,\cite{Gouveia-Neto}  without  the  need  of additional
dispersion compensation. This method exploits  the peculiarities of
high-order periodical soliton propagation in optical  fibers in the
anomalous dispersion regime, where efficient compression is obtained
at the output end of a fiber of properly chosen length. A signature
of this process  is  the appearance of a broad pedestal, which for
relatively long pulses ($>100$\,fs) can be efficiently suppressed
by  taking advantage  of  the birefringence induced by the intense
main pulse in the fiber itself\,\cite{Ahmed,Gong}.

Until  recently,  all  theoretical  and  experimental  studies  of
soliton compression were focused  on  normal  single-mode
fibers\,\cite{Agrawal},  which exhibit anomalous dispersion only for
wavelengths larger than $1\,\mu$m,  as  well as a relatively low
nonlinearity.  The  introduction  of  highly  nonlinear photonic
crystal fibers (PCFs) \cite{Russell} and photonics nanowires having
anomalous dispersion at visible and near infrared wavelengths has
shed a new light on this field, in particular in view of the
possibility  of applying  soliton compression  techniques  to  the
pulses  emitted by  today's  most  common ultrafast lasers, such as
Ti:sapphire and Cr:LISAF oscillators.

Specifically, the broad region of anomalous group  delay  dispersion
(which for photonic nanowires extends into the visible spectral
region) allows  for the  efficient  compression of pulses  in  the
$800$ \,nm  region,  as  recently demonstrated by  Foster  et
al.\,\cite{Foster},  who  experimentally  achieved pulse compression
of low-energy pulses from $70$\,fs  down  to  $6.8$\,fs  in  photonic
nanowires, and  theoretically  predicted  pulses  as  short  as  one
single optical cycle. Even if photonic nanowires  can  exhibit
flatter  dispersion profiles in the visible and near-infrared
regions than most PCFs, their  use still poses difficulties, mostly
due to  their  small  dimensions,  delicate construction, and
implementation. On the other hand, PCFs  with  core  sizes of around
$1.5\,\mu$m and a high nonlinearity are readily available. In a
recent work, Bessonov et al.\,\cite{Bessonov} theoretically
investigated  solitonic propagation and compression of high-energy
laser pulses in  hollow  photonic crystal fibers, where pulse
durations  down  to  $10$\,fs  at  wavelengths  of around $1\,\mu$m
were predicted.  However,  to  our  knowledge, soliton
self-compression of low-energy $800$\,nm ultrashort pulses in highly
nonlinear  PCFs and in the few-cycle regime has not been studied so
far.

In this work we investigate the possibility  of  efficient
self-compression of low-energy  laser  pulses  down  to  the
few-cycle  regime  using  a commercially available highly-nonlinear
PCF.

We perform a systematic numerical study, based on a generalized
nonlinear Schr\"odinger  equation that includes  higher-order
dispersion   terms, delayed Raman response and self-steepening, to
identify the  most  relevant parameters that determine  the
compression  limit.  A  detailed  comparison between different
approximations to nonlinear pulse propagation is made,  in order to
better isolate the detrimental effects that  more  strongly  affect
the compression process, which shows that pulses with durations a
short  as $4$\,fs can be directly  obtained  by  propagating  linearly
chirped or transform limited low-energy  pulses generated  from  a
typical ($30-100$\,fs) ultrafast Ti:sapphire oscillator in a PCF
without  the need of additional compression  methods,  which  may
result  in  novel  and compact devices for the generation of laser
pulses in the few-cycle regime.

We believe that this study will be very helpful for  predicting  the  output
of actual experiments based on this technique.

\section{Model description}
\begin{figure}[h]
 \centerline{\scalebox{0.55}{\includegraphics{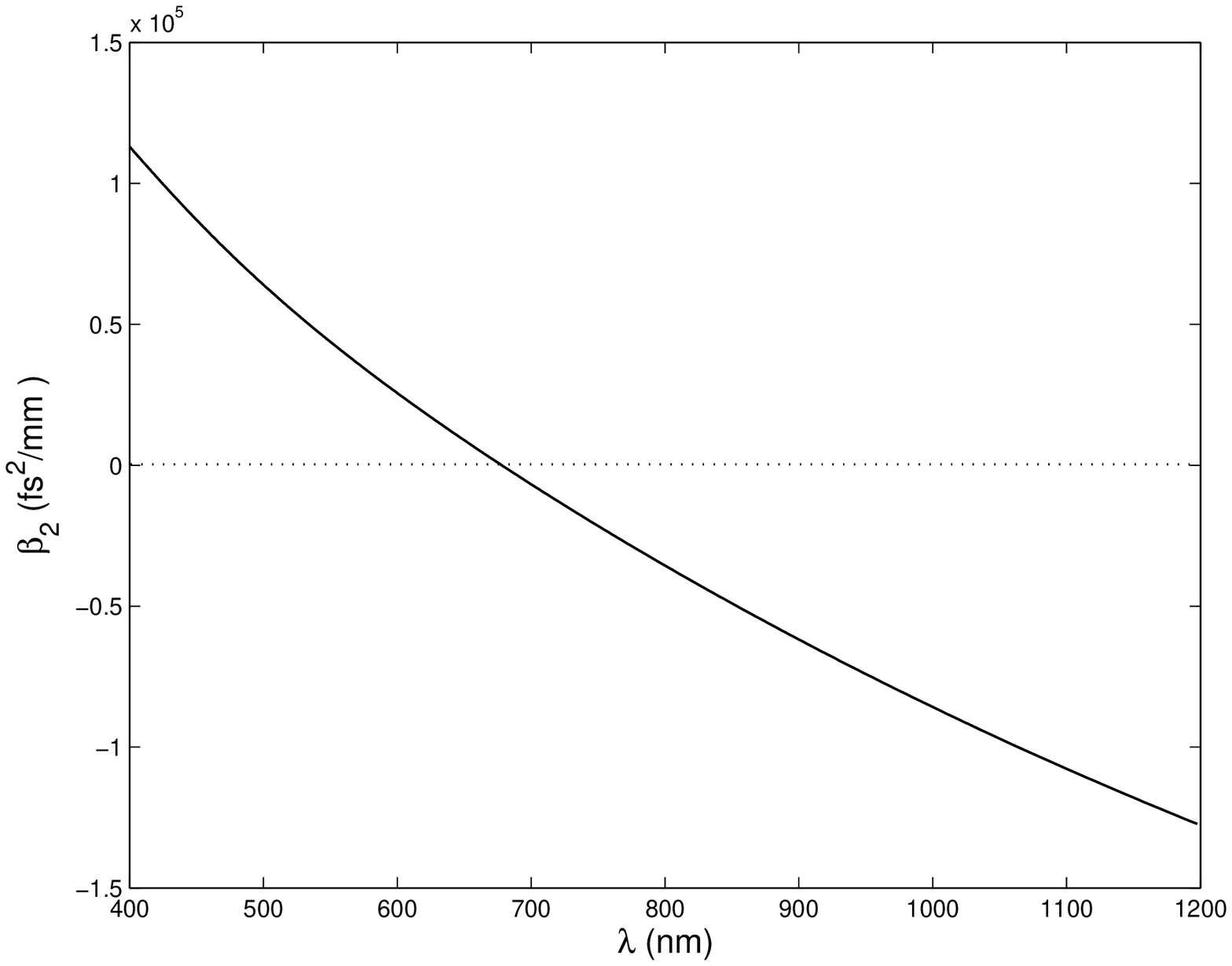}}}
 \caption{Group velocity dispersion of the highly-nonlinear PCF BlazePhotonics NL-1.6-670 (obtained from the manifacturer's data).}
 \label{figura1_sol_comp}
\end{figure}

In our model, an ultrafast laser pulse as directly generated from a
typical ($30-100$\,fs) commercially available Ti:sapphire laser
oscillator is coupled to a highly-nonlinear PCF of length $L$. The
propagation equation along the longitudinal fiber direction $z$ for
an electric field envelope $A(z,t)$ of central frequency $\omega_0$
and in the plane wave approximation is given by:
\begin{eqnarray}
\frac{\partial A(z,t)}{\partial z} &=& i\int_{-\infty}^{+\infty}
\beta(\omega) \tilde{A}(z,\omega)e^{-i\omega t} d
\omega\nonumber\\
& &+i B[A(z,t),t],
\label{prop_eq}
\end{eqnarray}
where $\tilde{A}(z,\omega)$ is the spectral field amplitude,
\begin{eqnarray}
\beta (\omega)&=&\sum_{n=2}^{+\infty} \beta_n(\omega_0) (\omega-\omega_0)^n
\label{beta}
\end{eqnarray}
 is the phase distortion, with $\beta_n(\omega)=\partial \beta(\omega)/\partial \omega$
 the $nth$-order dispersion, and
\begin{eqnarray}
B[A(z,t),t] &=& \gamma(1+\frac{i}{\omega_0}\frac{\partial}{\partial t})(A(z,t)\int_{0}^{+\infty}R(t')|A(z,t-t')|^2 dt')
\label{B}
\end{eqnarray}
is the nonlinear term  with $\gamma$ the nonlinearity coefficient
 and $R(t)$ the nonlinear response function. The functional form of $R(t)$ can be written as\,\cite{Agrawal}
\begin{eqnarray}
R(t)=(1-f_R)\delta(t)+f_R h_R(t),
\label{R}
\end{eqnarray}
where $f_R$ is the fractional contribution of the delayed Raman
response function $h_R(t)$, which can be expressed as:
\begin{eqnarray}
h_R(t)= \frac{\tau_1^2+\tau_2^2}{\tau_1\tau_2^2}exp(-t/\tau_2) \sin(t/\tau_1),
\label{h_R}
\end{eqnarray}
where $\tau_1$ and $\tau_2$ are two adjustable parameters. Numerical
values of $f_R=0.18$, $\tau_1=12.2$\,fs and $\tau_2=32$\,fs are
chosen to provide a good fit to the measured Raman gain spectrum of
fused silica\,\cite{Agrawal}. The Generalized Nonlinear
Schr\"odinger Equation (GNSE)\,(\ref{prop_eq}) includes both
self-steepening and the delayed Raman effect\,\cite{Agrawal} and
accurately describes pulse propagation of few nJ pulses in PCFs
down to the single single cycle regime in the framework of the
Slowly Evolving Wave Approximation (SEWA), which requires that the
envelope and its phase do not vary $significantly$ as the pulse
covers a distance equal to its central wavelength $\lambda_0=2 \pi
c/ \omega_0$\,\cite{Brabec1}. In the hypothesis of $negligible$
third-order dispersion ($\beta(\omega)\simeq 1/2 \beta_2(\omega_0)
(\omega-\omega_0)^2$), self-steepening, and delayed Raman response
($B\simeq \gamma A(z,t)|A(z,t)|^2$), equation\,(\ref{prop_eq})
reduces to the Nonlinear Schr\"odinger Equation (NSE), suitable for
describing pulse propagation in the picosecond or even
sub-picosecond temporal regime \,\cite{Agrawal}, provided that
third- and higher-order dispersion terms can be neglected and pulse
duration is significantly longer that the delayed Raman response of
the medium, $T_R=\int_0^{+\infty}t R(t) dt \approx 5$\,fs (for fused
silica). The fiber parameters are those  of a $commercial$
highly-nonlinear PCF (BlazePhotonics NL-1.6-670) already available
in our laboratory and which we plan to use in a forthcoming
experiment: $L=5$\,mm, $\gamma=139$\,W$^{-1}$\,mm$^{-1}$, and a group
velocity dispersion ($\beta_2$) profile as shown in
figure\,\ref{figura1_sol_comp}.

\section{Soliton-effect compression}

Soliton-effect compression relies on propagation properties of
optical solitons which are generated in nonlinear fibers in the
anomalous dispersion regime\,\cite{Agrawal}. In particular, solitons
of order $N \geq 2$ always change their shape periodically as they
propagate inside the fiber and in general experience an initial
pulse narrowing phase, which can be exploited to obtain pulse
compression once a proper fiber length is chosen\,\cite{Agrawal}.
The optimal fiber length $z_{opt}$ and the compression factor $F_c$
can be estimated from the following empirical relations obtained in
the framework of the NSE\,\cite{Agrawal}:
\begin{eqnarray}
F_c &\simeq& 4.1\,N\\
z_{opt} &\simeq& (\frac{0.32}{N}+\frac{1.1}{N^2})z_0,
\label{sol_eff_rel}
\end{eqnarray}
where $N=\sqrt{L_D/L_{NL}}$ is the soliton order, $L_D \simeq 0.321
\times T_{in}^2/\beta_2$ is the dispersion length, $L_{NL}=(\gamma
P_0)^{-1}$ is the nonlinear length, $z_0=\frac{\pi}{2} L_D$ is the
soliton period, $T_{in}$ is  the pulse initial
full-width-at-half-maximum (FWHM) duration, and $P_0$ is the peak
power of the initial pulse. In general, the compressed pulses
exhibit a broad pedestal whose origin is due to the fact that the
nonlinearity induces a linear chirp only over the central part of
the pulse, which is the only part that can be efficiently compressed
by the fiber anomalous group velocity dispersion \,\cite{Agrawal}.
The quality factor $Q_c$, defined as the fraction of the total
energy contained in the compressed pulse, is always less than unity
and scales as the inverse of the compression factor
$F_c$\,\cite{Agrawal}. Equations\,(\ref{sol_eff_rel}) predict an
indefinitely increasing compression factor for an initial fixed
pulse temporal profile with increasing peak power, once a proper
fiber length is chosen. However, equations\,(\ref{sol_eff_rel}) are
obtained by integrating the NSE, which fails to be valid in the few
femtosecond temporal regime, so the GNSE has to be used instead.

\section{Results and discussion}

\begin{figure}[h]
 \centerline{\scalebox{0.55}{\includegraphics[angle=270]{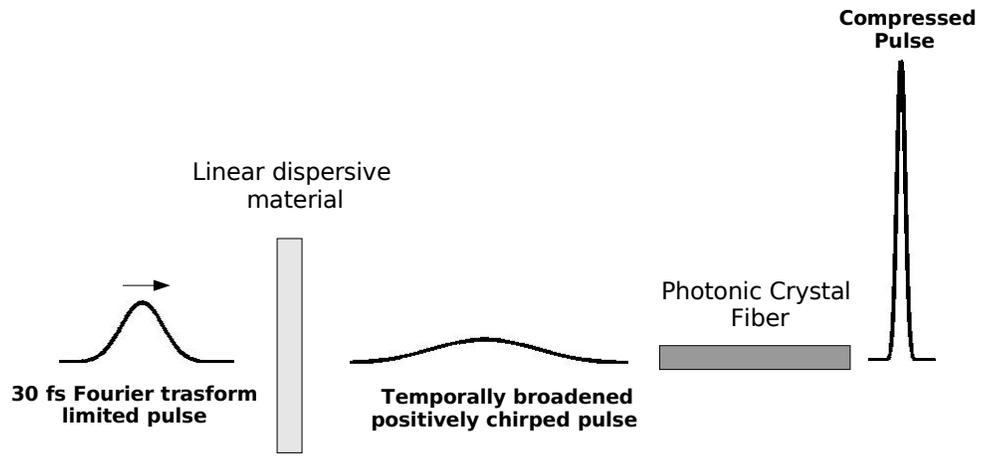}}}
 \caption{Scheme of the compression process.}
 \label{figura2_sol_comp}
\end{figure}

The best conditions for soliton pulse compression were obtained
starting from a $30$\,fs initial Fourier transform-limited (TL)
gaussian pulse with central wavelenght $\lambda_0=800$\,nm, which
can be generated from the ultrafast Ti:sapphire laser oscillator we
intend to use in our future
experiment\,\cite{Crespo2005,Tognetti2006}. Our numerical study
shows that a typical $30$\,fs few-nJ femtosecond pulse requires an
optimum fiber length considerably shorter than $5$\,mm, which
immediately poses a practical problem as it is very difficult to
obtain such short fiber lengths. However, this length can be
increased if the initial pulse duration is made larger for the same
pulse energy and we found that initial pulse durations of around $100$\,fs readily allow to overcome this difficulty.  Therefore, we chose
to introduce a positive linear dispersive material to temporally
broaden the initial $30$\,fs pulse so as to produce the optimal pulse
peak intensity  and temporal width at the fiber input (see
figure\,\ref{figura2_sol_comp}). We also found that this method
resulted in the same compressed pulse as if the spectral width
(hence temporal duration) of the initial TL pulse was varied
instead. The former method was nevertheless preferred to the latter,
as in principle it permits arbitrarily large extra-cavity pulse
stretching while retaining the initial pulse spectrum generated from
the laser oscillator, hence avoiding disturbances to the laser
parameters and mode-locking stability.

\begin{figure}[h]
 \centerline{\scalebox{0.85}{\includegraphics[angle=0]{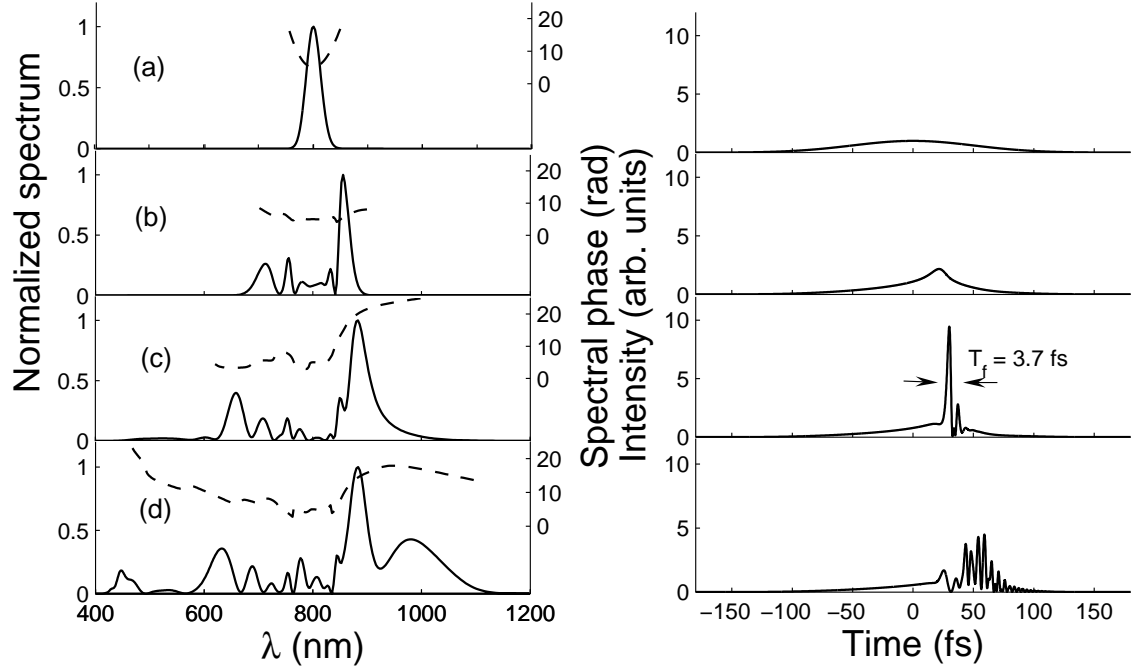}}}
 \caption{Spectra, spectral phases, and temporal profiles corresponding to (a) $z=0$, (b) $z=4$\,mm, (c) $z=5$\,mm, (d) $z=6$\,mm for a $30$\,fs laser pulse of initial energy $E=5\times 10^{-10}$\,J that has been temporally broadened to $119$\,fs.}
 \label{figura3_sol_comp}
\end{figure}

\begin{figure}[h]
 \centerline{\scalebox{0.8}{\includegraphics{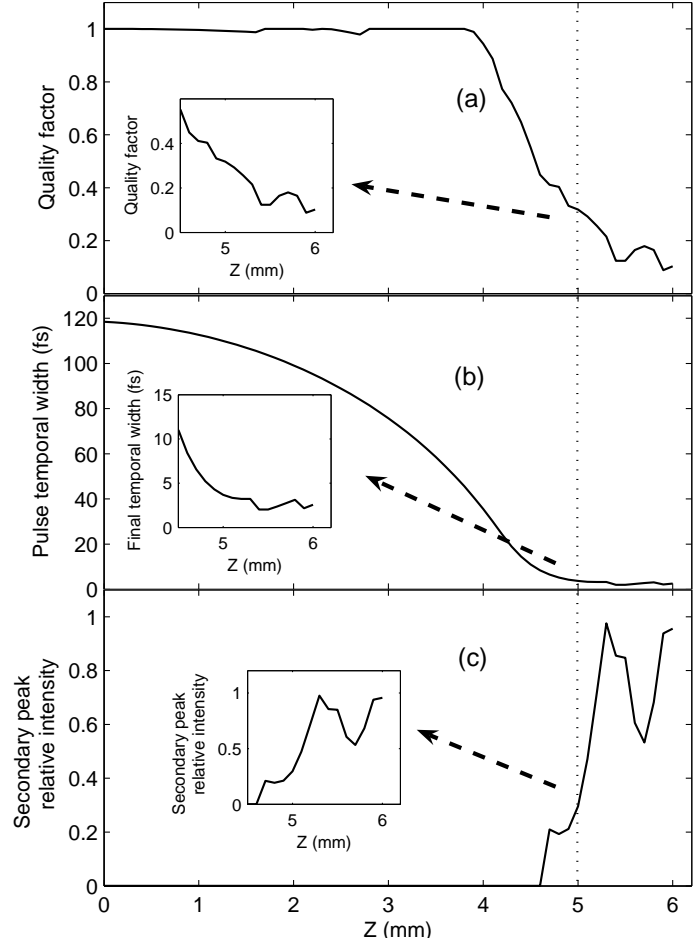}}}
 \caption{(a) Quality factor, (b) temporal width, and (c) relative intensity of the secondary peak as a function of propagation distance inside the PCF for the same initial pulse of figure\,\ref{figura3_sol_comp}. The dotted vertical line denotes the fiber length.}
 \label{figura4_sol_comp}
\end{figure}

\begin{figure}[h]
 \centerline{\scalebox{1.2}{\includegraphics[angle=0]{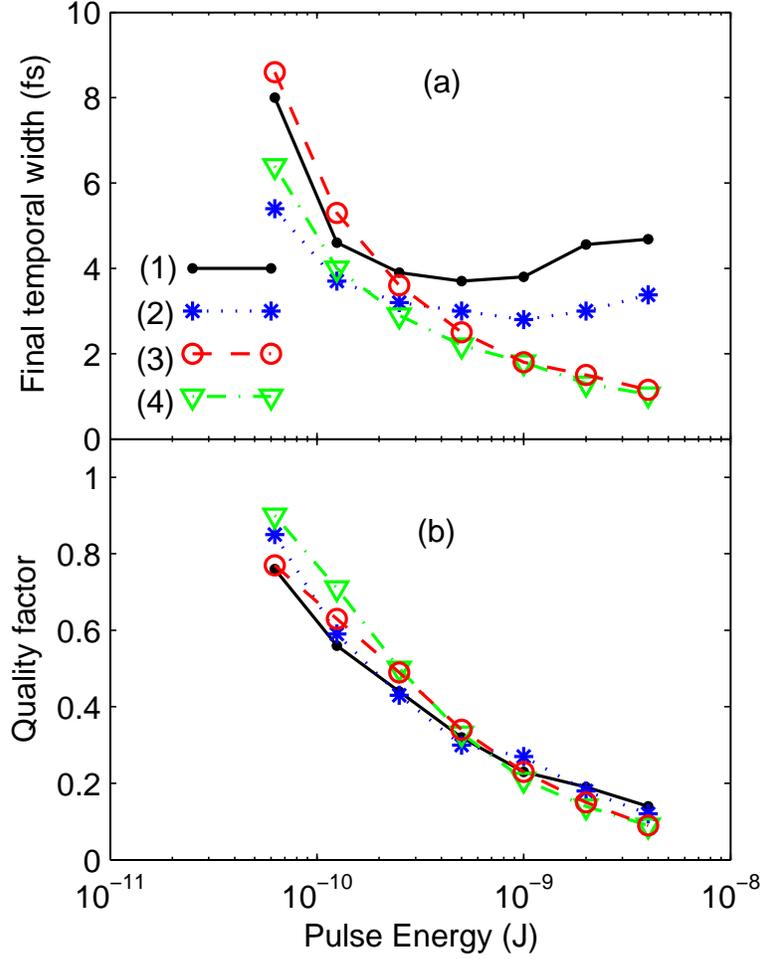}}}
 \caption{(Color online) (a) Minimum final temporal width and (b) quality factor as a function of the initial pulse energy obtained by integrating the GNSE (curves (1)), neglecting higher-order dispersion (curves (2)), neglecting self-steepening and delayed Raman response(curves (3)), and integrating the NSE (curves (4)). The reported values correspond to the narrowest obtainable pulse with a pre/post pulse  intensity lower than $0.3$ times the pulse peak value.}
 \label{figura5_sol_comp}
\end{figure}

\begin{figure}[h]
 \centerline{\scalebox{0.65}{\includegraphics[angle=0]{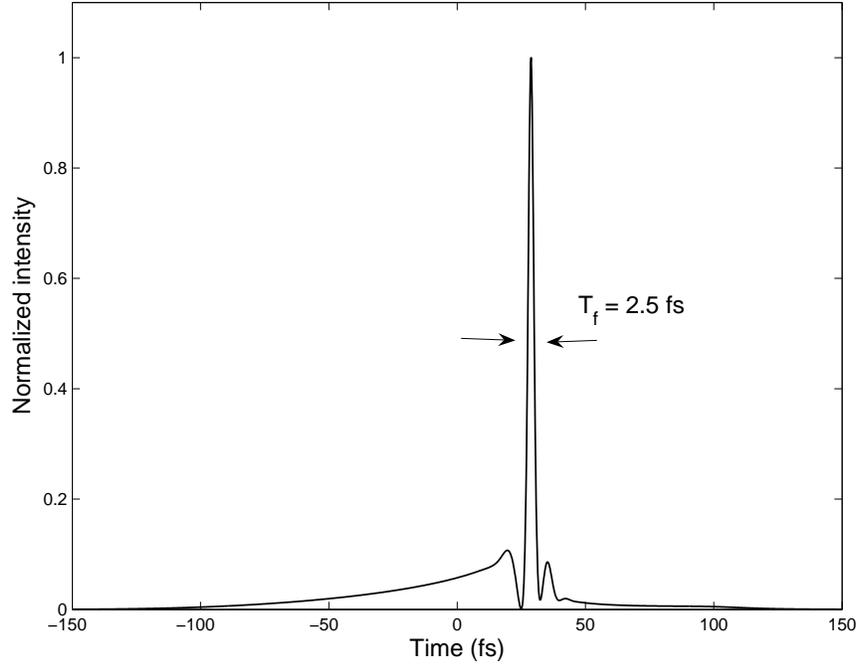}}}
 \caption{Single-cycle pulse obtained for the same initial pulse of figures\,\ref{figura3_sol_comp} and\,\ref{figura4_sol_comp}, in the hypothesis of a $completely$ flat dispersion profile, i.e. $\beta(\omega)\simeq 1/2 \beta_2(\omega_0) (\omega-\omega_0)^2$.}
 \label{figura6_sol_comp}
\end{figure}

Figure\,\ref{figura3_sol_comp}  shows the spectrum, the spectral
phase and the temporal profile of the pulse (a) at the fiber input,
and for fiber lengths of (b) $z=4$\,mm, (c) $z=5$\,mm, and (d)
$z=6$\,mm, assuming an initial pulse  energy $E=5\times 10^{-10}$\,J
and a pulse duration broadened to $119$\,fs in a normal positively
dispersive medium such as a piece of glass, so as to obtain an
optimized compressed pulse for $L=5$\,mm under the condition that
the most intense pre/post-pulse has an intensity lower than $0.3$
times the pulse peak value. It can be observed that the compression
process mostly acts upon the central part of the pulse temporal
profile, while maintaining  a broad uncompressed pedestal, until  a
pulse FHWM duration $T_f=3.7$\,fs and a quality factor $Q_c=0.32$
are reached. The central wavelength of the resulting broadened
spectrum does not deviate significantly from $800$\,nm, and so the
compressed pulse is less than two cycles in duration. For $z>5$\,mm
(see figure\,\ref{figura3_sol_comp}\,(d) and (e)) the pulse temporal
profile presents an increasing multi-peak and broadened structure.
The asymmetric spectral profile, larger in the $blue$ side, and the
steeper temporal trailing edge shown in
figure\,\ref{figura3_sol_comp} (c)  are those typical of
self-steepening, while Raman scattering is characterized by a shift
of the pulse spectrum towards lower frequencies, which is associated
with a temporal delay of the pulse\,\cite{Agrawal}. In
figure\,\ref{figura4_sol_comp} the spatial evolution of the temporal
pulse width, the quality factor and the relative intensity of the
largest secondary pre/post pulse are reported, showing how the pulse
compression process is always associated with a reduction in pulse
$quality$. The behavior shown in figures\,\ref{figura3_sol_comp}
and\,\ref{figura4_sol_comp} differs from the soliton pulse evolution
predicted by the NSE as the higher-order dispersion and nonlinear
effects included in the GNSE  destroy the periodical evolution
typical of high-order solitons. A systematical numerical study was
performed to identify which processes have the most detrimental
effect in the pulse compression process.
Figures\,\ref{figura5_sol_comp}\,(a) and(b) show the final temporal
width and the quality factor as a function of the pulse energy,
obtained by integrating the GNSE (curves (1)), neglecting
higher-order dispersion (curves (2)), neglecting self-steepening and delayed Raman response (curves (3)), and integrating the NSE (curves (4)). The
reported values correspond to the narrowest obtainable pulse with a
pre/post pulse relative intensity lower than $0.3$. It can be
noticed that higher-order dispersion terms have the most detrimental
effect in the compressed pulses, making curves (1) and (2) in
figure\,\ref{figura4_sol_comp} (a) go through a minimum value and
therefore deviating from the monotone behavior predicted by
relations\,(\ref{sol_eff_rel}) and confirmed by curves (3) and (4).
In this view the way of increasing the pulse compression effect
mostly relies on using a nonlinear fiber with the flattest possible
group velocity dispersion. These results are reminiscient of those
obtained for longer pulse durations in the $100$\,fs
range\,\cite{Chan}. In the ideal hypothesis of a completely flat
group velocity dispersion profile, for the same pulse energy and
temporal width of figure\,\ref{figura3_sol_comp}, a compressed
single-cycle pulse with $T_f=2.5$\,fs and $Q_c=0.34$ can be obtained
(see figure\,\ref{figura6_sol_comp}). The need of additional
dispersive polarization elements that introduce significant pulse
broadening prevents the direct application of standard nonlinear
birrefringe methods to suppress the observed broad pedestal found in
soliton compressed pulses. The pedestal could nevertheless be
partially suppressed by simply focusing the pulses in a thin
($100-300\mu$m), low-dispersion near infrared low-bandpass filter,
which can act as an efficient saturable absorber for femtosecond
pulses\,\cite{Jiang}.

\section{Conclusions}
In conclusion, we numerically demonstrate the feasibility of
efficient soliton compression of transform-limited or linearly
chirped ultrashort laser pulses down to the sub-2-cycle regime using
a standard Ti:sapphire oscillator and a $5$-mm long commercially
available PCF. An optimized $3.7$\,fs pulse can be obtained from an
initial ultrashort laser pulse centered at $800$\,nm, with duration
in the $100-$fs range and an energy of $0.5$\,nJ, while longer pulses
can also be compressed at slightly higher pulse energies of a few
nJ. We identify high-order dispersion as the most relevant
detrimental factor in few-cycle soliton compression, showing that
single-cycle pulses with $2.5$\,fs can be obtained for the ideal case
of a PCF with a completely flat dispersion profile. We believe that
this technique could be the basis for novel, compact and efficient
sources of few-cycle laser pulses based on a standard ultrafast
oscillator coupled to a properly chosen PCF and a simple pulse
cleaner, which could have a great impact in the scientific
community.

\section{acknowledgments}
This work was partly supported by FCT Grant No.
POCTI/FIS/48709/2002, Portuguese Ministry of Science, co-financed by FEDER.


\newpage

\newpage

\newpage

\newpage

\newpage

\newpage

\end{document}